\documentclass[prl,twocolumn,superscriptaddress,showpacs]{revtex4}
\usepackage{amsfonts,amssymb,amsbsy,bm,graphicx}

\newcommand{\Tr}{\mathop{\mathrm{Tr}}\nolimits}

\begin{document}

\title{Quantum reconstruction of an intense
polarization squeezed optical state}

\author{Ch.~Marquardt}
\affiliation{Institute of Optics, Information and Photonics
(Max Planck Research Group),
University of Erlangen-Nuremberg,
G\"{u}nther-Scharowsky Stra{\ss}e 1, Building 24,
91058 Erlangen, Germany}

\author{J.~Heersink}
\affiliation{Institute of Optics, Information and Photonics
(Max Planck Research Group),
University of Erlangen-Nuremberg,
G\"{u}nther-Scharowsky Stra{\ss}e 1, Building 24,
91058 Erlangen, Germany}

\author{R.~Dong}
\affiliation{Institute of Optics, Information and Photonics
(Max Planck Research Group),
University of Erlangen-Nuremberg,
G\"{u}nther-Scharowsky Stra{\ss}e 1, Building 24,
91058 Erlangen, Germany}

\author{M.~V.~Chekhova}
\affiliation{Department of Physics,
M. V. Lomonosov Moscow State University,
119992~Moscow, Russia}

\author{A.~B.~Klimov}
\affiliation{Departamento de F\'{\i}sica,
Universidad de Guadalajara,
44420~Guadalajara, Jalisco, Mexico}

\author{L.~L.~S\'anchez-Soto}
\affiliation{Institute of Optics, Information and Photonics
(Max Planck Research Group),
University of Erlangen-Nuremberg,
G\"{u}nther-Scharowsky Stra{\ss}e 1, Building 24,
91058 Erlangen, Germany}

\author{U.~L.~Andersen}
\affiliation{Institute of Optics, Information and Photonics
(Max Planck Research Group),
University of Erlangen-Nuremberg,
G\"{u}nther-Scharowsky Stra{\ss}e 1, Building 24,
91058 Erlangen, Germany}
\affiliation{Department of Physics,
Technical University of Denmark,
Building 309, 2800 Lyngby, Denmark}

\author{G.~Leuchs}
\affiliation{Institute of Optics, Information and Photonics
(Max Planck Research Group),
University of Erlangen-Nuremberg,
G\"{u}nther-Scharowsky Stra{\ss}e 1, Building 24,
91058 Erlangen, Germany}
\date{\today}

\begin{abstract}
We perform a reconstruction of the polarization sector
of the density matrix of an intense polarization squeezed
beam starting from a complete set of Stokes measurements.
By using an appropriate quasidistribution, we map this
onto the Poincar\'e space providing a full quantum
mechanical characterization of the measured polarization
state.
\end{abstract}

\maketitle

Efficient methods of quantum-state reconstruction
are of the greatest relevance for quantum optics.
Indeed, they are invaluable for verifying and
retrieving information. Since the first theoretical
proposals~\cite{firsttheory} and the pioneer
experiments determining the quantum state of a
light field~\cite{firstexp}, this discipline has
witnessed significant growth~\cite{Jarda}.
Laboratory demonstrations of state tomography
are numerous and span a broad range of physical
systems, including molecules~\cite{DWM1995},
ions~\cite{LMK1996}, atoms~\cite{KPM1997},
spins~\cite{spin}, and entangled photon
pairs~\cite{WJE1999}.

Any reliable quantum tomographical scheme
requires three key ingredients~\cite{HMR2006}:
the availability of a tomographically complete
measurement, a suitable representation of the
quantum state, and a robust algorithm to invert
the experimental data. Whenever these conditions
are not met, the reconstruction becomes difficult,
if not impossible. This is the case for the
polarization of light, despite the fact that
many recent experiments in quantum optics
have been performed using polarization states.
The origin of these problems can be traced back
to the fact that the characterization of the
polarization state in terms of the total density
operator is superfluous because it contains not
only polarization information. This redundancy
can be easily handled for low number of
photons, but becomes a significant hurdle
for highly excited states. An adequate
solution has been found only recently:
it suffices to reconstruct only of a subset
of the density matrix. This subset has been termed
the ``polarization sector"~\cite{RMF2000} (or the
polarization density operator~\cite{Kar2005}) since
its knowledge allows for a complete characterization
of the polarization state~\cite{KM2004}.

The purpose of this Letter is to report on the first
theoretical and experimental reconstruction of intense
polarization states. Specifically, we focus on the case
of intense squeezed states to confirm how, even in this
bright limit, they still preserve fingerprints of very
strong nonclassical behavior.

We begin by briefly recalling some background material.
We assume a two-mode field that is fully described
by two complex amplitude operators, denoted by $\hat{a}_H$
and $\hat{a}_V$, where the subscripts $H$ and $V$ indicate
horizontally and vertically polarized modes, respectively.
The commutation relations of these operators are standard:
$[\hat{a}_j, \hat{a}_k^\dagger ] = \delta_{jk}$, with
$j, k \in \{H, V \}$. The description of the polarization
structure is greatly simplified if we use the Schwinger
representation
\begin{eqnarray}
\label{Stokop}
& \hat{J}_1 = \frac{1}{2} ( \hat{a}^\dagger_H \hat{a}_V +
\hat{a}^\dagger_V \hat{a}_H ) \, ,
\qquad
\hat{J}_2 =  \frac{i}{2} ( \hat{a}_H \hat{a}^\dagger_V -
\hat{a}^\dagger_H \hat{a}_V ) \, ,
& \nonumber \\
& \hat{J}_3 =  \frac{1}{2} ( \hat{a}^\dagger_H \hat{a}_H -
\hat{a}^\dagger_V \hat{a}_V ) \, , &
\end{eqnarray}
together with the total photon number $\hat{N} =
\hat{a}^\dagger_H \hat{a}_H + \hat{a}^\dagger_V \hat{a}_V$.
These operators coincide, up to a factor 1/2, with the Stokes
operators, whose average values are precisely the classical
Stokes parameters. One immediately finds that $\hat{\mathbf{J}}
= (\hat{J}_1, \hat{J}_2, \hat{J}_3)$ satisfies the commutation
relations distinctive of the su(2) algebra: $[\hat{J}_1, \hat{J_2}]
= i \hat{J}_3,$ and cyclic permutations. This noncommutability
precludes the simultaneous  precise measurement of the physical
quantities they represent.

The Hilbert space $\mathcal{H}$ describing the polarization
structure of these fields has a convenient orthonormal
basis in the form of the Fock states for both polarization
modes, namely $|n_H, n_V \rangle$. However, it is advantageous
to use the basis $|J, m \rangle$ of common eigenstates of
$\hat{J}^2$ and $\hat{J}_3$. Since $J = N/2$, this can be
accomplished just by relabeling the Fock basis as $|J, m \rangle
\equiv | n_H = J+m, n_V = J-m \rangle$. Here, for fixed $J$
(i.e., fixed $N$), $m$ runs from $-J$ to $J$ and these states
span a $(2J+1)$-dimensional subspace wherein $\hat{\mathbf{J}}$
acts in the standard way. Since any polarization observable
has a block-diagonal form in this basis, it seems appropriate
to define the polarization density operator as
\begin{equation}
\hat{\varrho}  =  \bigoplus_{J=0}^\infty \hat{\varrho}_J =
\sum_{J=0}^\infty \ \sum_{m, m^\prime = - J}^{J}
\varrho^J_{m m^\prime} \ |J, m \rangle \langle J, m^\prime | \, .
\end{equation}

\begin{figure}
\includegraphics[width=0.95\columnwidth]{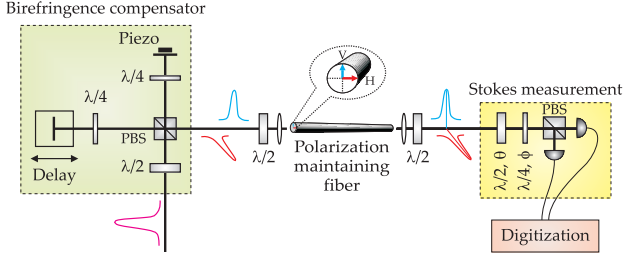}
\caption{(Color online) Setup for efficient polarization
squeezing generation and the corresponding Stokes
measurement apparatus.}
\end{figure}

For a reconstruction of the quantum state one first has
to extract the required data from the tomographical
measurements. The overall scheme of our experimental
setup is illustrated in Fig.~1. The field to be
characterized is analyzed using a general
polarization measurement apparatus consisting of
a half-wave plate $(\lambda /2 ,\theta)$ followed
by a quarter-wave plate ($\lambda /4, \phi$) and
a polarizing beam splitter (PBS). The wave plates
transform the input polarization allowing the
measurement of different Stokes parameters by
the projection onto the PBS basis $| J, m \rangle$.
The PBS outputs are measured directly using detectors
with custom-made InGaAs photodiodes (98\% quantum
efficiency at DC) and a low-pass filter ($\leq$ 40~MHz)
to avoid AC saturation due to the laser repetition rate.
The RF currents of the photodetectors were mixed
with an electronic local oscillator at 17.5~MHz
and digitized with an analog/digital converter
at $10^7$~samples per second with a 16~bit
resolution and 10~times oversampling. The
quantum state we measure is defined by the
resolution bandwidth of 1~MHz at the 17.5~MHz
sideband relative to the 200~THz carrier. Ten
digitized sample corresponds to the photocurrent
at this sideband generated by photons
impinging on the photodiode for 1~$\mu$s. This
photodetection can be modelled by the positive
operator-valued measure (POVM)~\cite{agliati05}
\begin{equation}
\hat{\Pi}_{m}^J = |J, m \rangle \langle J, m | \, ,
\end{equation}
so that $w_{m}^J =  \Tr ( \hat{\varrho}  \hat{\Pi}_{m}^J )$
is the probability of detecting $n_H = J+m$ photons in the
horizontal mode and simultaneously $n_V= J-m$ photons in
the vertical one. When the total number of photons $2J$
is not measured and only the difference $2m$ is observed,
the POVM is
\begin{equation}
\hat{\Pi}_{m} = \sum_{J = |m|}^\infty
|J, m \rangle \langle J, m | \, .
\end{equation}

The wave plates in the measurement perform linear
polarization transformations. These can be described
in terms of $\hat{J}_2$, which generates rotations
about the direction of propagation, and $\hat{J}_3$,
which generates phase shifts between the modes. In
other words, their action is represented by $\hat{\mathcal{R}}
(\mathbf{n}) = e^{ i \theta \hat{J}_2} \,
e^{ i \phi  \hat{J}_3}$, where $\mathbf{n} =
(\cos \phi \sin \theta, \sin \phi \sin \theta,
\cos \theta)$  is a  unit vector given by the spherical
angles $(\theta, \phi)$. The experimental histograms
recorded for each $\mathbf{n}$ then correspond to
the tomographic probabilities
\begin{equation}
\label{cojo}
w^J_m (\mathbf{n} ) =  \Tr [ \hat{\varrho} \,
\hat{\Pi}_m^J (\mathbf{n}) ] =
{}_{\mathbf{n}}\!\langle J, m | \hat{\varrho}
| J, m \rangle_{\mathbf{n}} \, ,
\end{equation}
where $ \hat{\Pi}_{m}^J ( \mathbf{n} ) =  \hat{\mathcal{R}}
(\mathbf{n}) \, \hat{\Pi}_m^J \, \hat{\mathcal{R}}^\dagger
(\mathbf{n})$ and $|J, m \rangle_{\mathbf{n}}$ is the
eigenstate of $\mathbf{n} \cdot \hat{\mathbf{J}}$ relative
to the eigenvalue $m$, which is precisely a SU(2) coherent
state~\cite{Per1986}. The final tomogram reads
\begin{equation}
w_m (\mathbf{n} )  =  \Tr [ \hat{\varrho} \
\hat{\Pi}_m (\mathbf{n}) ] = \sum_{J = |m|}^\infty \
_{\mathbf{n}}\!\langle J, m | \hat{\varrho}
|J, m \rangle_{\mathbf{n}} \, .
\end{equation}

The reconstruction in each $(2J+1)$-dimensional invariant
subspace can be now carried out exactly since it is
essentially equivalent to a spin $J$~\cite{spintom}.
In fact, after some calculations one finds that the
tomograms can be represented in the following compact
form
\begin{equation}
w^J_m (\mathbf{n} ) = \frac{1}{2 \pi} \int_0^{2 \pi}
d\omega \ \Tr \left ( \hat{\varrho}_J \ e^{i \omega \,
\mathbf{n} \cdot \hat{\mathbf{J}}} \right ) \,
e^{- i m \omega} \, ,
\end{equation}
which is precisely the Fourier transform of the
characteristic function of the observable $\mathbf{n}
\cdot \hat{\mathbf{J}}$.

Inverting this expression, one obtains
\begin{eqnarray}
\hat{\varrho}_J & =  & \frac{2J+1}{4\pi^2}
\int_0^{2\pi} d \omega \ \sin^2 \left ( \frac{\omega}{2} \right )
\nonumber \\
& \times & \int_{\mathcal{S}_2} d \mathbf{n} \ e^{-i\omega \,
\mathbf{n} \cdot \hat{\mathbf{J}}}
\sum_{m=-J}^{J} w_m^J (\mathbf{n}) \, e^{i m \omega } \, ,
\end{eqnarray}
where the integration over the solid angle $d\mathbf{n} =
\sin \theta \, d\theta  d\phi$ extends over the unit sphere
$\mathcal{S}_2$. By summing over all the invariant subspaces
$J$, the density matrix can be reconstructed as follows
\begin{equation}
\label{uf}
\hat{\varrho}  =   \sum_{m=-\infty }^{\infty}
\int_{\mathcal{S}_2} d\mathbf{n} \ w_{m}(\mathbf{n} ) \,
\mathcal{K} (m - \mathbf{n} \cdot \hat{\mathbf{J}} ) \, ,
\end{equation}
where the kernel $\mathcal{K} (x) $ is
\begin{equation}
\mathcal{K} ( x ) = \frac{ 2 J +1}{4 \pi^2}
\int_{0}^{2\pi}d\omega \ \sin^2 \left ( \frac{\omega}{2} \right ) \,
e^{-i \omega  x} \, .
\end{equation}
We stress the appealing analogy of Eq.~(\ref{uf}) with the
more widely known formula for the reconstruction of the
density matrix of a single-mode radiation field from the
homodyne tomograms of the rotated quadratures~\cite{Jarda}.

From the exact solution (\ref{uf}), one can calculate
any polarization quasidistribution~\cite{WSU2}. From
a computational point of view reconstructing the SU(2)
$Q$ function turns out to be the simplest, since in each
invariant subspace it reduces to
\begin{equation}
Q(J,\mathbf{n} ) =
{}_{\mathbf{n}}\!\langle J, m |  \hat{\varrho}
|J, m \rangle_{\mathbf{n}} \, ,
\end{equation}
so, in view of the form (\ref{cojo}), it is especially
suited for our purposes. The evaluation of the  Wigner
function can also be carried out, although with additional
effort. Nevertheless, we do not expect these two
quasidistributions to differ notably for the states
we study here. As a consequence, we only need to evaluate
the matrix elements of the kernel $\mathcal{K} (m -
\mathbf{n}^\prime \cdot \hat{\mathbf{J}})$. This
can be accomplished using several equivalent techniques,
although the most direct way to proceed is to note that
\begin{eqnarray}
\label{evalK}
& \displaystyle
{}_{\mathbf{n}}\!\langle J, m | \mathcal{K} (m -
\mathbf{n}^\prime \cdot \hat{\mathbf{J}} )
| J, m \rangle_{\mathbf{n}}  =  \frac{2J+1}{4 \pi^2}
\int_{0}^{2\pi} d\omega \
 \sin^2 \left ( \frac{\omega}{2} \right ) &
\nonumber \\
& \displaystyle
\times \ e^{im \omega} \left [ \cos \left (
\frac{\omega}{2} \right ) - i \sin \left (
\frac{\omega}{2} \right ) \cos \chi \right ]^{2J} \, ,
&
\end{eqnarray}
where $\cos \chi = \mathbf{n} \cdot \mathbf{n}^\prime$.
In the limit of $J \gg~1$ the integral in (\ref{evalK})
reduces to $d^2 \delta (x) /dx^2$ evaluated at
$x = m - J \, \mathbf{n} \cdot \mathbf{n}^\prime$. Since
$m$ can be taken as a quasicontinuous variable, we can
integrate by parts
\begin{equation}
Q (J, \mathbf{n}) = \frac{2J+1}{4\pi^2}
\int_{-\infty }^{\infty} \! dm
\int_{\mathcal{S}_2} \!  d\mathbf{n}^\prime \,
\frac{d^2 w_{m} ( \mathbf{n})}{dm^2}  \,
\delta (m - J \, \mathbf{n} \cdot \mathbf{n}^\prime ) \, .
\end{equation}
Thus, in the limit of high photon numbers the reconstruction
turns out to be equivalent to an inverse 3D Radon
transform~\cite{Dea1983} of the measured tomograms,
which greatly simplifies the numerical evaluation of
$Q (J, \mathbf{n})$.

To test this theory, we  prepared polarization squeezed
states by exploiting the Kerr nonlinearity experienced by
ultrashort laser pulses in optical fibers~\cite{heersink05}.
Our experimental setup shown in Fig.~1 uses a Cr$^{4+}$:YAG
laser emitting 140~fs FWHM pulses at 1497~nm with a repetition
rate of 163~MHz. Using the two polarization axes of a 13.2~m
birefringent fiber (3M FS-PM-7811, 5.6~$\mu$m mode-field diameter),
two quadrature squeezed states are simultaneously and
independently generated. The emerging pulses' intensities
are set to be identical and they are aligned to temporally
overlap. The fiber's polarization axes exhibit a strong
birefringence (beat length 1.67~mm) that must be compensated.
To minimize post-fiber losses, we precompensate the pulses
in an unbalanced Michelson-like interferometer that introduces
a tunable delay between the polarizations~\cite{heersink03}.
A small part (0.1~\%) of the fiber output serves as the input
to a control loop to maintain a $\pi/2$ relative phase between
the exiting pulses, producing a circularly polarized beam.

Since the Kerr effect is photon-number conserving, the
amplitude fluctuations of the two individual modes
$H$ and $V$ are at the shot-noise level. This was
checked using a coherent beam from the laser and employing
balanced detection. The average output power from the fiber
was 13 mW which, with the bandwidth definition of our
quantum state, corresponds to an average number of
photons of $10^{11}$ per 1 $\mu$s. The Kerr effect in
fused silica generates squeezing up to some Terahertz,
making the choice of the sideband in principle arbitrary.
The wave plates in the measurement apparatus were rotated
by motorized stages. These scanned one quarter of the
Poincar\'e sphere in 64 steps for $\phi$ ($0^\circ -
22.5^\circ$) and 65 steps for $\theta$ ($0^\circ -
45^\circ$), a measurement of which took over 5 hours.
The rest of the data can be deduced from symmetry. For
each pair of angles, the photocurrent noise of both
detectors after the PBS was simultaneously sampled
$2.9 \times 10^6$~times. Noise statistics of the
difference of the two detectors' photocurrents were
acquired in histograms with 2048 bins, resulting in
the tomograms $w_m (\mathbf{n})$. The optical intensity
incident on both detectors was recorded as well.

\begin{figure}
\includegraphics[width=0.90\columnwidth]{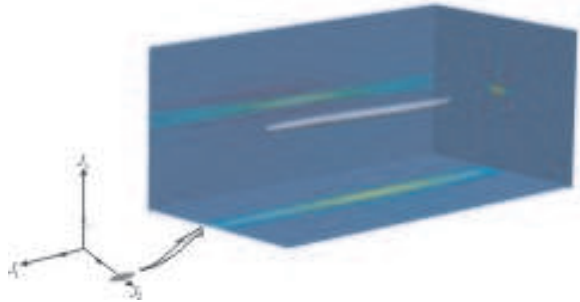}
\caption{(Color online) Three-dimensional isocontour
surface plot of the measured $Q (J, \mathbf{n})$
function for a polarization squeezed state. In
the inset, we show the projections over the
coordinate planes passing through the origin
of the ellipsoid. The projection on the $J_1$-$J_3$
plane shows typical artifacts from the Radon
transform.}
\end{figure}

To reconstruct the $Q$ function we performed a 3D
inverse Radon transform. In Fig.~2 we show the result
of the reconstruction for a polarization squeezed
state. Here an isocontour surface of the 3D space
(or Poincar\'e space) of $Q (J, \mathbf{n})$ is seen,
as well as the projections of this surface. The
ellipsoidal shape of the polarization squeezed
state is clearly visible. The antisqueezed
direction of the ellipsoid is dominated by excess noise
stemming largely from Guided Acoustic Brillouin Scattering
(GAWBS), which is characteristic for squeezed states generated
in optical fibers. Note that this reconstruction of an intense
Kerr squeezed state is very hard with conventional homodyne
detection techniques for quadrature variables due to the
high intensities. The projection of the ellipsoid on the
plane $J_2$-$J_3$ results in an ellipse (and not a circle)
because of imperfect polarization contrast in the measurement
setup. As the classical excitation of the state is in the
$J_2$ direction, one expects to reach the shot-noise limit
in this projection. After the interference at the PBS
all intensity impinges on only one detector. However,
with limited polarization contrast some parts of
the $>$ 25 dB excess noise from the dark output port
can be mixed into the intense beam. Acting as a homodyne
measurement this will be visible in the noise of the detector.

In Fig. 3 we compare the isocontour surface plots of a
coherent and a polarization squeezed state for the value
corresponding to half width at half maximum. The size of
the contours agree with the 6.2 $\pm$ 0.3 dB squeezing
that was directly measured with a spectrum analyzer.

\begin{figure}
\includegraphics[width=0.90\columnwidth]{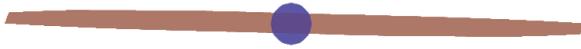}
\caption{(Color online) Meridional sections of the
isocontour surface plots of the $Q$ function for
a coherent state (blue) and a polarization squeezed
state (red).}
\end{figure}

By summing over all the values of $J$, we can obtain
the total $Q (\mathbf{n})$, which is a probability
distribution over the Poincar\'e unit sphere and is
properly normalized. In Fig.~4 we have plotted such
a function for the squeezed state. As the state has
a large excitation of the order of $10^{11}$ photons
and the angles of the distribution on the unit sphere
are small, the spherical coordinates can be treated
like Cartesian coordinates in the vicinity of the
classical point and we present a zoomed version of the
surface of the sphere. Again the excess phase noise
of the squeezed state is visible. The oscillations
with negativities near the main peak are characteristic
of artifacts arising from the inverse Radon transformation.
These artifacts are more visible for sharp or elongated
structures. To improve the reconstruction one would
have to go to more sophisticated reconstructions,
e.g., maximum likelihood methods~\cite{Jarda}.

\begin{figure}[h]
\includegraphics[width=0.80\columnwidth]{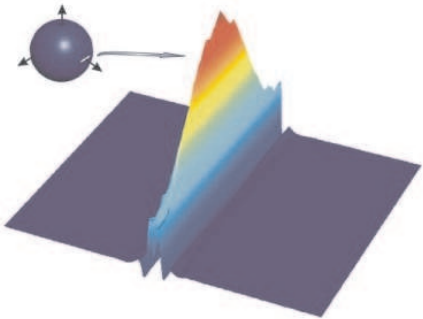}
\caption{(Color online) Probability distribution
over the Poincar\'e unit sphere for a polarization
squeezed state obtained over all invariant subspaces.
The distribution is strongly concentrated at the
classical mean value, so we show a zoomed version.}
\end{figure}

In summary, we have presented an exact inversion formula
for quantum polarization states and derived a simplified
version for high-intensity states. The reconstruction of
an intense polarization squeezed state, formed by the
Kerr effect in an optical fiber, was performed. Interesting
future investigations include the comparison with the
maximum likelihood method and the reconstruction of
nonclassical polarization states with lower intensity.

We thank V. P. Karassiov for useful discussions and
C.~M\"{u}ller for technical assistance. Financial
support from the EU (COVAQIAL No FP6-511004),
CONACyT (Grant 45704), and DGI (Grant FIS2005-0671)
is gratefully acknowledged. M. V. Ch. was supported
by DFG (Grants 436 RUS 17-75-05 and 436 RUS 17-76-06).

\end{document}